\documentclass[a4paper,11pt]{article}
\usepackage{jheppub} % for details on the use of the package, please see the JINST-author-manual
\usepackage{hyzsyntax}

\arxivnumber{2404.16097} % if you have one

\title{\boldmath Generalized Symmetries in String-constructed QFTs via K-theory}

\author{Hao Y. Zhang}
\affiliation{Kavli Institute for the Physics and Mathematics of the Universe (WPI),\\
University of Tokyo, Kashiwa, Chiba 277-8583, Japan}

\emailAdd{hao.zhang@ipmu.jp}

\abstract{We propose that generalized symmetries in some string-constructed QFTs are given by K-theory. We thus have \textit{even-form} and \textit{odd-form} symmetries determined by $K_N(\p X)$, the twisted K-theory as D-brane charges on the asymptotic boundary $\p X$ of internal geometry $X$ with twist class $N$. For these QFTs, ``\textit{$p$-form symmetries}" are no longer separately well-defined for individual $p$, but are instead mixed together. We discuss 6D ADE-type (2,0) SCFTs and some 6d (1,0) LSTs as examples and demonstrate their twisted K-theoretic symmetries, and we checked them to be compatible with T-duality. We further point out, through explicit examples, that K-theory leads to symmetry extensions that cannot be detected by cohomology for Type II string theory on certain orbifolds of $\mathbb{C}^3$ and $\mathbb{C}^4$. We also discuss the implications of these results in the dual brane descriptions.}

\begin{document}
\maketitle
\flushbottom

\section{Introduction}

Generalized symmetries \cite{Gaiotto:2014kfa} are symmetries that act on not only point objects but also extended objects. The initial generalization is $p$-form symmetry acting on spacetime $p$-dimensional operators or spatially $p$-dimensional excitations. Such symmetries are defined to be the topological operators that, after crossing a charged object, exerting phase shift on them. Later, a further generalization is higher-group symmetries \cite{Tachikawa:2017gyf,Cordova:2018cvg,Benini:2018reh} which can be seen as a non-trivial extension of several $p$-form symmetries. Another direction of generalizations are non-invertible symmetries \cite{Bhardwaj:2017xup,Chang:2018iay}, namely the topological operators satisfy non-invertible fusion rules, and they may map genuine operators to non-genuine ones.

The usual assumption is that $p$-form symmetries are separately well-defined for each $p$, and charged objects with different dimensionalities are also distinct. Even for $n$-group symmetries, their definitions requires the notion of individual $p$-form symmetries to start with. However, in this paper, we propose yet another possible generalization of the notion of symmetry, by considering symmetry operators and charged objects which are intrinsically not restricted to a specific dimensionality. Indeed, in strongly coupled QFTs constructed from string theory, heavy defects of different dimensionality can sometimes be charged under the same symmetry group, since the string-theoretic object used to construct them can be related to the dynamical process of tachyon condensation. Therefore, the symmetry operators themselves no longer wrap cycles w.r.t. ordinary homology of the spacetime manifold, but equivalently wrap cycles w.r.t. cretain generalized homology of the spacetime which captures topological charges that are invariant under tachyon condensation. In addition, the algebraic object characterizing such symmetry should also be given by suitable generalized (co)homology of the \textit{internal manifold}. 

We take a preliminary step by focusing on a family of SCFTs/LSTs constructed from type II string theory, where we argue that the generalized symmetries are described by K-theory.
\footnote{Here we remain agnostic about more general cases, e.g., when the brane used to construct the generalized global symmetry is coupled to NS-NS sector field.} 

 K-theory \cite{atiyah1962vector,bott1963lectures} can be understood as as classifying the topological configuration of a pair of complex vector bundles $(E, F)$ on a manifold $M$, modulo the simultaneous addition of a third complex vector bundle $(E, F) \cong (E \oplus H, F \oplus H)$. K-theory  was identified to classify the RR-charges of D-branes \cite{Minasian:1997mm,Witten:1998cd} and configurations of RR fields \cite{Moore:1999gb}, where $(E, F)$ can be understood as the gauge bundle on D9 brane and anti-D9 brane stacks, respectively, and modding out the equivalence relation of simultaneously adding (thus removing) $H$ physically corresponds to tachyon condensation. 
   
 In QFTs engineered from type II string theory, our key evidence for K-theoretic symmetries comes by imposing the consistency under string duality. Namely, we insist that the generalized global symmetry in the compactified theories must be compatible with T-duality. Moreover, we take the viewpoint that in such string-constructed QFTs, generalized symmetry operators also admit natural top-down constructions (see \cite{GarciaEtxebarria:2022vzq,Apruzzi:2022rei} in the context of holography, and \cite{Heckman:2022muc} in the context of geometric engineering). This consistency requirement on pairs of type-II string theories under T-duality leads to a dramatic consequence, that we can no longer describe generalized symmetries as branes wrapping \textit{individual} \textit{ordinary} homological cycles. Instead, we must take the twisted K-theory charge of branes and reduce it on twisted K-theory cocycles of the boundary of internal manifolds. The fact that brane charges and RR-fields are described in K-theory was discovered in the 90s in \cite{Minasian:1997mm,Witten:1998cd,Moore:1999gb}. But in the literature of generalized symmetries, the usage of K-theory remains at a conceptual level \cite{GarciaEtxebarria:2022vzq,Apruzzi:2022rei,Heckman:2022muc,Bah:2023ymy} or in specific examples. See \cite{Heckman:2022xgu} where twisted K-theory is introduced in an example to obtain a non-trivial homology in 6D $\calN = (2,0)$ and thus 4D $\calN = 4$ SYM, \cite{Garcia-Etxebarria:2017crf,Torres:2024sbl} proposing to use a KO-theoretic field for anomaly cancellation in 8D, and \cite{GarciaEtxebarria:2024fuk} for discussions of symmetry TFT involving K-theory. The process of open-string tachyon condensation is built into the K-theory description of D-brane charges, whose counterpart for topological operators is carefully analyzed in \cite{Bah:2023ymy}. \footnote{The terminology of K-theoretic symmetry also appears in the review \cite{Loaiza-Brito:2024hvx}.} 

Before closing this section, we compare our notion of mixing $p$-form symmetries by twisted K-theory with that of \textit{higher-group symmetries} which also involves mixing of symmetries at different form degrees \cite{Tachikawa:2017gyf,Cordova:2018cvg,Benini:2018reh}. For example, a 2-group symmetry mixes two symmetries of \textit{adjacent} degree, typically 0-form and 1-form. Physically, it involves the phenomenon that the F-move of codimension-1 symmetry operators produces a codimension-2 symmetry operator. On the other hand, K-theoretic symmetries only mix a set of $q$-form symmetries where $q$ are either all even or all odd. It treats all symmetry operators of different dimensionalities by modding out an equivalence relation among them, the latter is induced by tachyon condensation from string theory. Therefore, we can view K-theoretic symmetry as a specific type of higher-group symmetries, which only involves even/odd form symmetries while skipping odd/even form symmetries. \footnote{From the top-down point of view, 2-group symmetry admit string-theory origin \cite{DelZotto:2022fnw,DelZotto:2022joo,Cvetic:2022imb} in terms of M-theory, which is related to the Mayer-Vietoris sequence on the asymptotic boundary $\p M$ of the internal manifold $M$. It would be interesting to explore the relation between 2-group symmetry via M-theory compactifications and symmetry in type II string compactifications described by K-theory, as connected by T-duality.}

The rest of this paper is organized as follows. In Section \ref{sec:HtoK}, we give backgrounds for defect groups and higher-form symmetries in geometric engineering, and explain that the same answer cannot be reproduced in the dual brane constructions. We conclude that characterizing generalized symmetries with ordinary cohomology is incompatible with T-duality. In Section \ref{sec:K-theory}, motivated by the K-theory classification of D-brane charges, we propose that the generalized symmetries in type II compactifications should be described by twisted K-theory, whose compatibility with T-duality is well-established. We illustrate our proposal by giving examples of 6D (2,0) SCFTs and some (1,0) LSTs that admit no type I or heterotic constructions. In some (2,0) cases, we point out that the K-theoretic symmetries have been explicitly reproduced from the worldsheet. In Section \ref{sec:TachyonCondensation}, we proceed to the physical implication of such a K-theoretic description of generalized symmetries in terms of symmetry operators and charged states. In section \ref{sec:c3c4}, we provide more examples of K-theory theoretical symmetry in internal manifolds, and we point out important cases where the K-theoretical symmetry explicitly differes from cohomology in $\bbC^4$ orbifolds theories, and further discussed their implications. \footnote{We relegate some technical material to the Supplemental Material, including a detailed treatment of K-theoretic global symmetries in 6D (2,0) of DE-type, and identification of twisted K-theoretic symmetry from the string worldsheet.}

\section{Review and Motivation}
\label{sec:HtoK}

\textbf{Generalized symmetries and defect group} Generalized symmetries are usually defined in a ``standard" Quantum Field Theory, one that has a scalar-valued partition function, known as an \textit{absolute} QFT. When studying a QFT obtained from string compactication, however, one does not directly obtain an absolute QFT. Instead, we usually get a \textit{relative quantum field theory} to start with \cite{Freed:2012bs}, from which we need to pick a maximal set of commuting fluxes on the asymptotic boundary to obtain an absolute QFT, and discuss the generalized symmetries of the latter.  The best known example of a relative QFT is 2D WZW model (whose partition function is called conformal blocks) that has to live at the boundary of 3D Chern-Simon theory \cite{Witten:1988hf}. 

In general, a relative QFT cannot be consistently defined on its own, but has to live at the boundary of a topological QFT (TQFT) in one dimension higher. Relative QFT has to admit a set of flux operators arise from reduction of dual fields on torsional homological cycles in the internal geometry. There flux operators obey Heisenberg flux non-commutativity, so they cannot consistently act on a 1-dimensional partition function. Therefore, such relative field theories only have a partition vector as opposed to a (scalar-valued) partition function. 

The higher-form symmetries of the absolute QFT comes from the \textit{defect group} of the underlying relative QFT,
which contains non-dynamical (``heavy") objects \cite{DelZotto:2015isa}. Therefore, we need to understand the defect group as a first step. Defect group is defined as a sum over the groups of charges of extended defects of $k$ spacetime dimensions:
 \begin{equation}
     \bbD = \bigoplus_k \bbD_k.
 \end{equation}
On the defect group $\bbD$, there is a fractional valued Dirac pairing
 \begin{equation}
     \langle \cdot, \cdot \rangle: \bbD \times \bbD \rightarrow \bbQ/\bbZ
 \end{equation}
 encoding the Heisenberg flux non-commutativity of these non-dynamical defect operators. Whenever possible, we could then pick a maximal commuting subset of flux operators (so that the Dirac pairing among these operators become \textit{integral}) to make it an absolute QFT. See \cite{Aharony:2013hda} for systemic study of all possible absolute theories in 4D $\calN = 4$ SYM, and \cite{DelZotto:2015isa,Lawrie:2023tdz} for the general procedure of producing absolute theories from the defect group with more examples in 6D.

 Passing from a relative QFT to an absolute QFT requires us to pick a maximal subset of commuting operators.\footnote{Such an operation is sometimes called ``picking a polarization". From the bulk perspective, it can be understood as picking a topological boundary condition for the bulk TQFT.} At the level of defect group, this requires picking a polarization (i.e., a Lagrangian subgroup of the defect group $L \subset \bbD$ such that $|L|^2 = |\bbD|$) and the Dirac pairing on $\bbD$ vanishes on when restricted on $L$.
 
 As a concrete example, in 4D we have $SU(N)$ gauge theory with heavy defect given by a fundamental Wilson line $W_{2\pi/N}$, and $PSU(N) = SU(N)/\bbZ_N$ gauge theory with heavy defect given by a fundamental 't Hooft line $H_{2\pi/N}$. The label $1/N$ is a manifestation of 't Hooft Screening \cite{tHooft:1977nqb,tHooft:1979rtg}, namely N copies of a fundamental (quark-anti-quark) Wilson line can be screened by a dynamical gluon. The self-pairings of these lines vanish $\langle W_{2\pi/N}, W_{2\pi/N} \rangle, \langle H_{2\pi/N}, H_{2\pi/N} \rangle = 0 \in \bbZ$, so they are indeed well-defined absolute theories. However, there is an underlying relative theory labelled by the $\ksu(N)$ Lie algebra. It contains both the fundamental Wilson lines and fundamental 't Hooft lines, and their mutual Dirac pairing is fractional: $\langle H_{2\pi/N}, H_{2\pi/N} \rangle = 1/N$, so the corresponding flux operators do not commute. After picking a maximal set of commuting operators, one would end up with the absolute theories of $SU(N)$ and $PSU(N)$ (or other cases for specific $N$). In the language of defect group, $\bbD = \bbZ_{N}^{(e)} \oplus \bbZ_{N}^{(m)}$, and picking the $SU(N)$ (or $PSU(N)$ absolute theory corresponds to choosing the Langrangian algebra to be $L = \bbZ_{N}^{(m)}$ (or $\bbZ_{N}^{(e)}$). \footnote{Such a choice of maximal subset of commuting operators (i.e., a choice of Lagrangian subgroup $L \subset \bbD$) is always possible in 4D, since the Dirac pairing in 4D is anti-symmetric so the pairing of any element with itself always vanish in $\bbQ/\bbZ$. On the other hand, this is not always possible in 6D where such pairing instead becomes symmetric, e.g., the 6D (2,0) theory of type $A_1$ and defect group $\bbZ_2$ does not lead to any absolute theory.}

 % \hyz{old content:} , where it follows from the original argument of 't Hooft that $N$ units of such fundamental Wilson line could be screened by the gauge boson that carries the adjoint representation. In a word, we would need defect group because it is a piece of data in a relative QFT that is the best analog of higher-form symmetry in an absolute QFT.

\textbf{Defect group via geometric engineering} From geometric engineering of QFT via string theory, the defects with non-trivial change under the defect group is given by wrapping D-branes on relative homological cycles of the internal manifold $\gamma_{rel}$ representing an element in $H_*(X_{D-d}, \partial X_{D-d})$, a relative homological $n$-cycles in the internal space $X_{D-d}$ with boundary $\p X_{D-d}$. Here the total spacetime is $D = 10$ dimensional, which takes the form of $M_{d} \times X_{D-d}$ with $M_{d}$ the macroscopic spacetime and $X_{D-d}$ the internal space. On such a relative cycle with infinite volume, it is possible to wrap a D-brane to produce a heavy / non-dynamical defect, whose mass is formally infinite. However, such a construction might also produce elements with trivial charges in the defect groups, and we need to mod out these trivial-charge objects to compute the defect group itself. This happens for branes wrapping a relative cycle in $\gamma_{rel} \in H_*(X_{D-d}, \partial X_{D-d})$ that can be pulled into the bulk of the internal geometry, i.e., it is (the image of) a cycle as an element in $\gamma \in H_*(X_{D-d})$. These cycles $\gamma$ have vanishing volumes, and thus D-branes wrapped on them produces dynamical particles which screens the charge of the non-dynamical defects, reproducing the aforementioned 't Hooft Screening argument from string theory. Therefore, geometric engineering gives us the following expression of the defect group \cite{DelZotto:2015isa,Albertini:2020mdx,Morrison:2020ool}:\begin{equation}
    \mathbb{D} = \bigoplus_k \bbD^{(k)}, \quad \bbD^{(k)} = \bigoplus_{n, p} \frac{H_n(X_{D-d}, \partial X_{D-d})}{H_n(X_{D-d})} \ \ (\text{p-branes and n-cycles s.t. }k+n=p+1),
\end{equation} where we sum over all available branes in string theory (i.e., D$p$ branes for $p$ even in IIA and $p$ odd in IIB). Here for each $n$, we are including a summand of $H_n(X_{D-d}, \p X_{D-d})$, the charge of non-dynamical defects, modulo $H_n(X_{D-d})$, the 't Hooft screening effect of dynamical objects on which some multiple of a unit-charged defect can terminate. In the context of string compactification, the procedure of picking a polarization has been explained in type IIB string compactification in \cite{DelZotto:2015isa,GarciaEtxebarria:2019caf,Gukov:2020btk,Lawrie:2023tdz}, explaining the choice of $L \subset \bbD$ in the context of geometric engineering and illustrated with 4D and 6D examples.

\begin{figure}
    \centering
    \includegraphics[width=0.9\linewidth]{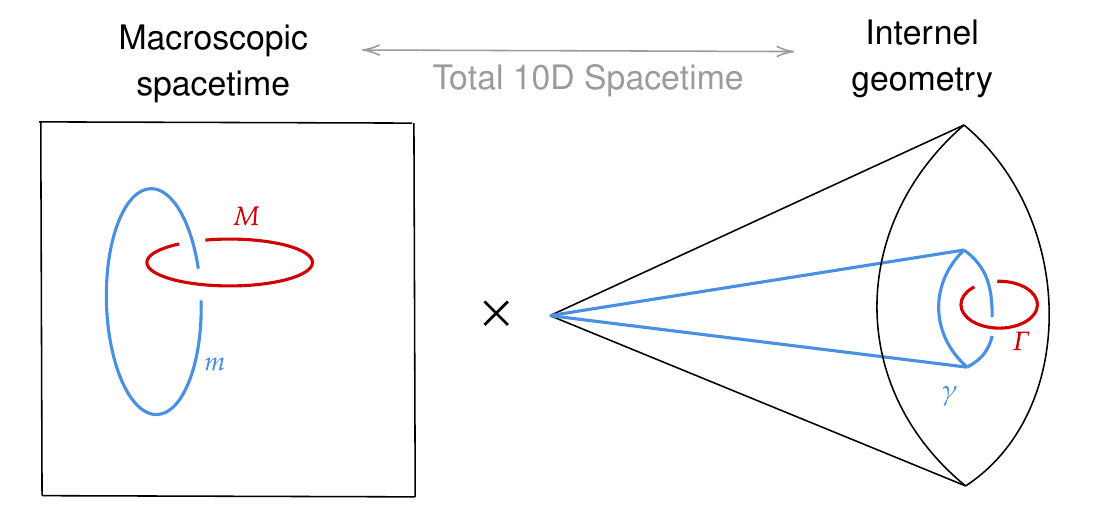}
    \caption{Compactification of the 10D string theory on a internal geometry (right) taken to be conifold singularity gives a effective QFT in the remaining macroscopic dimensions. In the internal space, the effective QFT is localized at the tip of the cone. BLUE: wrapping a D-brane on a \textit{relative} topological cycle $\gamma_{rel} \in H_*(X_{D-d}, \partial X_{D-d})$ to get a heavy defect $m$; RED: wrapping a D-brane on a topological cycle $\Gamma \in H_*(\p X_{D-d})$ to get a topological cycle $M$. The linking of $M$ and $m$ in the macroscopic dimensions can be reproduced by the Dirac pairing of the pair of D-branes in the total 10D spacetime.}
    \label{fig:braneboundary}
\end{figure}

\textbf{Symmetry operator perspective} Up till now, we have explained the constructions of the heavy defects that are charged under the defect group, but the corresponding symmetry operators remain implicit in the above construction (see \cite{GarciaEtxebarria:2022jky,Apruzzi:2022rei,Heckman:2022muc}). In fact, one could also directly examine the property of the symmetry operators, making the analysis of higher-form symmetries more explicit. Such symmetry operators can be constructed by having branes wrapping cycles representing elements in $H_*(\p X_{D-d})$, see figure \ref{fig:braneboundary} (adapted from \cite{Heckman:2022muc}). The fact that $\p X_{D-d}$ being infinitely far away from the tip of the conifold singularity (where the strongly-coupled QFT resides) makes the resulting extended operator decouple from any dynamics and thus become purely topological. The charged operator will thus be obtained by wrapping the \textit{magnetic dual} D-brane (D$(d-p-4)$ brane is dual D$p$ brane) on the linking relative homological cycle, reproducing the linking in the macroscopic dimensions from the internal geometry perspective.
We learn from algebra topology that $H_*(\p X_{D-d}) = \frac{H_n(X_{D-d}, \partial X_{D-d})}{H_n(X_{D-d})}$, so the symmetry operator perspective directly detects the defect group, circumventing the step of screening heavily defects. \footnote{We remark that the starting point of this discussion involves a topological operator in a relative QFT, which has not been carefully formulated yet. Such a formulation would potentially require extending the topological operator into the bulk TQFT, which would be an interesting direction to pursue in the future.}

\subsection{T-duality Rules Out Ordinary (co)Homology}

Having reviewed the standard argument of describing generalized symmetries and defect groups by \textit{ordinary} (co)homology, we now give motivation on why we want to revise these arguments using string duality (T-duality to be specific). 

For QFTs admitting any type II string constructions, such constructions always come in T-dual pairs: a type IIA construction on one manifold, and a type IIB construction on a T-dual manifold (or a Mirror dual manifold in general). As a specific set of cases, a pure geometric internal space (i.e., not having net flux of RR fields) is generically dual to a brane configuration, therefore focusing on either pure geometric constructions or pure brane constructions would be inadequate for a complete understanding incorporating T-duality. \footnote{In principle, we should also study heterotic and type I compactifications for completeness, but we defer those to future studies.} 

For the moment, we focus on type II compactification to 6D such as 6D (2,0) SCFTs (which descend to 4D $\calN = 4$ SYM by $T^2$ compactification) and some 6D (1,0) LSTs. By insisting that the generalized symmetry be manifestly compatible with T-duality, we will see that the formulation of generalized symmetries in terms of ordinary (co)homology no longer holds. Even though our general consideration holds for arbitrary dimensions, we would like to illustrate our idea by using 6D examples, where the T-duality is best understood, while leave other dimensionalities for future studies.

Notice that the IIA construction of 6D (2,0) theories has no geometric singularity, so the boundary of the internal geometry has cohomology $H^*(S^3) = \{\bbZ, 0, 0, \bbZ\}$. Therefore, there are no torsional (co)cycles that could produce this $\bbZ_N$ 2-form defect group as in the dual IIB side, nor can any torsional 2-form symmetry be detected when the relative theory descends to an absolute theory. One then realizes that D-brane charge in such a type IIA background takes value in twisted K-theory \cite{Moore:1999gb}. However, to see how far we can go without using K-theory, we attempt to describe the genearlized symmetries by twisted cohomology, with the twist given by the flux of the 2-form NS-NS field B with flux $\int H_3 = N_{NS_5}$. We will see that using twisted cohomology will allow us to superficially match the pair of groups describing the symmetries on both sides, but we could not reconcile the discrepancy between the meaning behind these two Abelian groups without finally uplifting to K-theory.

% [Maybe put this somewhere below] since it can be represented by differential forms with a modified cocycle condition.

\textbf{Twisted (Co)homology and Its Limitations} In the presence of non-trivial RR flux, the gauge transformation rule of RR potential $C_p$ (and thus their flux $G_{p+1}$) needs to be modified as:
\begin{align}
    \begin{split}
        C_p &\rightarrow C_p + d\Lambda_{p-1} - H \wedge \Lambda_{p-3} ( p \geq 3) \\
        G_{p+1} &\rightarrow G_{p+1} + H \wedge d \Lambda_{p-3},
    \end{split}
\end{align}
so that $G_{p+1}$ is no longer gauge-invariant. One can cancel this gauge variation by adding a compensating term% that cancels the $H \wedge d \Lambda_{p-3}$, which is well-known to be
\begin{equation}
    F_{p+1} = G_{p+1} - H \wedge C_{p-2} = d C_{p} - H \wedge C_{p-2}. \label{eqn:H_shifted_F}
\end{equation}
However, this gauge-invariant $F_{p+1}$ is no longer co-closed, but instead:
\begin{equation}
    dF_{p+1} = H \wedge F_{p-1}.
\end{equation}
thus $F_{p+1}$ is no longer a representative of ordinary (de Rham) cohomology of the spacetime, because $d$ is no longer restricted to a particular degree. However, we can still get a modified version of cohomology by considering the formal sum of all these gauge-invariant field strengths. Take type IIA for example:
\begin{equation}
    F_{\mathrm{even}} = F_2 + F_4 + F_6 + F_8 + F_{10}.
\end{equation}
This $F_{\mathrm{even}}$ is a representative of a $(d-H)$-twisted de Rham cohomology with a modified cocycle condition:\begin{equation}
(d-H) F_{even} = 0,\ \ [F_{even}] \in H_{N}(\p X),
\end{equation}
where $N = [H] \in H^3(\p X, \mathbb{Z})$ is the cohomology class of $H_3$ viewed as the twist class, and $\p X$ asymptotic boundary of the internal manifold $X$.\footnote{For 6D (2,0) theories of $D$-type, we also need to introduce a different $\bbZ_2$-valued twist, which can be found in the Appendix \ref{apdx:6D_DEtype}.} The $H_3$-twisted cohomology on $S^3$ with twist class $N = k \xi_3$ can be computed by definition ($\xi_3$ generates $H_3(S^3) = \bbZ$):\begin{equation}
H^*_{k\xi_3}(S^3) = \{\bbZ_N, 0\} \label{eqn:H_twist_S3},
\end{equation}so we attempt to construct generalized symmetry by reducing brane charges on this ``twisted" cocycle. This produces a torsional symmetry $\mathbb{Z}_N$, which coincides with the $\bbZ_N$ ``defect group" obtained from type IIB compactified on $\mathbb{C}^2/\bbZ_N$ as groups.

However, even though both type IIA and type IIB string theories each produce a $\bbZ_N$-valued symmetry, we should not ignore the mismatch on their physical meanings: the twisted homology only has two elements for any $X$, so the notion of dimensionality of the cycles does not make sense unless we reduce it modulo 2. Therefore, the $\bbZ_N$ ``\textit{2-form} defect group" on the IIB side is fundamentally different from the $\bbZ_N$ ``\textit{even-form} defect group" on the IIA side. In general, for dual pairs between pure geometric constructions and brane constructions involving NS5 branes, such a mismatch will persist as a fundamental issue if we were to continue understand the symmetry by twisted homology. Thus at this point, is fairly natural to further promote twisted cohomology to twisted K-theory, which will end up solving the above issue. 

\section{Full Treatment: Twisted K-theory} \label{sec:K-theory}

Now, we make use of the well-known fact that RR charges and RR fields are classified by the twisted K-theory of the spacetime manifold. From this we propose that, the generalized symmetry of the QFT coming from type II compactification is governed by twisted K-theory of asymptotic boundary $\p X$ of internal manifold $X$, promoting the result in \cite{Heckman:2022muc} to generalized cohomology. % and that of the external spacetime $M$,

K-theory (topological K-theory to be specific) is defined as the equivalence class of a pair of vector bundles $(E, E')$ modulo the simultaneous direct sum of a third bundle $F$:\begin{equation}
    (E, E') \cong (E \oplus F, E' \oplus F),    
\end{equation}
in string theory, $E, E'$ are interpreted as the gauge bundles on D-branes and anti-D-branes, respectively. This defines the $K$-group of degree 0, denoted as $K^0$. The equivalence relation involving $F$ captures the dynamical process of open string tachyon condensation, with the tachyonic modes living on open strings from D-brane to anti-D-brane. In string-engineered QFTs, such tachyon condensations underlie the fusion rules of symmetry operators from branes wrapping ordinary homological cycles \cite{Bah:2023ymy}.

Another important group $K^1$ (which is the same as $K^{-1}$ by Bott periodicity), can be specified as $K^{1}(X) = K(\Sigma X)$, where $\Sigma X$ is the suspension of $X$. We know there are two elements one needs to specify for K-theory of complex bundle due to the bott periodicity
\begin{equation}
    K^*(M) = K^{*+2}(M),
\end{equation}
with the period element given by the canonical line bundle over $\bbC \bbP^1 \cong S^2$.

On top of the (bundle characterization of) ordinary K-theory we just introduced, twisted K-theory imposes extra structure on the pair of bundles, and sometimes one needs to formally think of both $E$ and $E'$ as infinite-dimensional vector bundles, and to define twisted K-theory using analytic objects. See \cite{atiyah2004twisted} for a version of definition (known as ``algebraic K-theory" in mathematics), which involves projective unitary operators, Fredholm operators, and $C^*$-algebras on a Hilbert space. Possible twisting are classified by an element in
\begin{equation}
    (a_1, x_3) \in H^1(X; \bbZ_2) \times H^3(X, \bbZ),
\end{equation}
and we denote the resulting twisted K-theory on the topological space $X$ as:
\begin{equation}
    ^{a_1}K_{x_3}(X).
\end{equation}

Here $x_3 = k \xi_3 \in H^3(X, \bbZ)$ is the twist associated with $k$ units of NS-NS $H_3$ flux (with $\xi_3$ a generator of $H^3(X, \bbZ)$, for $X$ a 3-manifold, $H^3(X, \bbZ) = \bbZ$), and $H^1(X; \bbZ_2)$ captures the effect of an orientifold that ``flips the pair of bundles" when going around an element representing the fundamental group of $X$. Here the explanation needs to be taken with a grain of salt, since the the definition of $(E, F) \sim (E \oplus H, F \oplus H)$ does not always have a clear analog when the twist takes a non-trivial value in $H^3(X, \bbZ)$.  Only for $x_3$ torsional, it is possible to find finite-dimensional twisted vector bundles \cite{atiyah2004twisted}. But when $\xi_3$ is a free element, we would need to morally replace $E, F$ with infinite-dimensional vector bundles. 

However, the algebraic K-theory definition mentioned above still incorporates the most general twist which has both a degree-1 component and a degree-3 component. There, the twist of K-theory is phrased as a Dixmier-Douady invariant \cite{BSMF_1963__91__227_0} (see \cite{schochet2009dixmier} for an accessible introduction in English) of the $C^*$ algebra bundle on the manifold $X$. We refer the interested reader to \cite{Mathai:2000iw}, a concise paper with definition and computation of $K_N^*(S^3)$ as an example. 

Unlike ordinary cohomology which admits a $\bbZ$-valued degree, twisted K-theory only admits a $\bbZ_2$-valued degree. So we only get to talk about ``even-form" and ``odd-form" symmetries. As we have seen, RR fields of different degrees can transform into each other via gauge transformation. On the other hand, only the K-theory class $x$ of the total spacetime $X_{\text{tot}}$ is the physical degree of freedom. Its image under the Chern character is represented by the formal sum of RR field strengths $F_{p+1}$. Concretely,
\begin{align}
    \begin{array}{c}
    IIA: \mathrm{ch}(x) = F_2 + F_4 + F_6 + F_8 + F_{10}, \ x \in K^0(X_{\text{tot}}),\\
    IIB: \mathrm{ch}(x) = F_1 + F_3 + F_5 + F_7 + F_9,\ x \in K^1(X_{\text{tot}}). \nonumber
    \end{array}
\end{align}

The object charged under the formal sum is the collection of all BPS branes. D-branes in IIB are charged under $K^0_{N}(X_{\text{tot}})$ by consider D9 branes and D9-$\overline{\text{D9}}$ pairs, and then D-branes in IIA are classified by $K^1_{\tilde{N}}(\tilde{X}_{\text{tot}})$ that is obtained via T-duality \cite{Minasian:1997mm,Witten:1998cd}.

\paragraph{K-theory of Spacetime and Internal Space}

It is instructive to examine the K\"{u}nneth formula of decomposing K-theory in the asymptotic boundary of the total spacetime $\p X_{\text{tot}} =M \times \p X$, with $M$ is the external spacetime and $X$ the internal manifold with asymptotic boundary $\p X$. As a reminder, $K$ has a feature of Bott periodicity $K^* \cong K^{*+2}$ which distinguish it from an ordinary cohomology theory. Therefore, to specify $K$ theory of a space $M$, one only need to determine two elements, $K^0(M)$ and $K^1(M)$.

The formula can be found in \cite{atiyah1962vector}, but we first examine the special case when $M$ has no torsion:   
\begin{small}
\begin{align}
    \begin{array}{c}
    K^0(X_{\text{tot}}) = K^0(M) \otimes K^0(\p X) \oplus K^1(M) \otimes K^1(\p X), \\
    K^1(X_{\text{tot}}) = K^0(M) \otimes K^1(\p X) \oplus K^1(M) \otimes K^0(\p X).
    \end{array}
\end{align}
\end{small}
When $M$ has torsion, mathematically $K^i(X_{\text{tot}})$ will be an extension of the right-hand side over a direct sum of $\mathrm{Tor}(K^i(M), K^j(\p X))$ terms for certain sets of $(i, j)$. 

In general, we expect to take the \textit{right-hand side} of the above equation as the \textit{definition} of even-form and odd-form symmetries, so there is no mixing between K-theory of $M$ and that of $\p X$. The last ingredient we need is to include the topological twist, which we now introduce. 

\textit{Twisted K-theory and Topological T-duality} When we have a (trivial or non-trivial) twist $x_3 \in H^3(X_{\text{tot}}; \bbZ)$ which decomposes as $x_3 = y_0 \cup z_3$ for $y_0 \in H^0(M; \bbZ)$ and $z_3 \in H^3(\p X, \bbZ)$. It is natural to consider the following form of K-theoretic symmetries
\begin{small}
\begin{align}
    \begin{array}{cc}
    \text{IIB} & K^0(M) \otimes K^0_{z_3}(\p X) \oplus K^1(M) \otimes K^1_{z_3}(\p X), \\
    \text{IIA} & K^0(M) \otimes K^1_{z_3}(\p X) \oplus K^1(M) \otimes K^0_{z_3}(\p X), \\
    \end{array} \label{eqn:decomposition}
\end{align}
\end{small}
where we need the twisted K-theory of $\p X$, but untwisted K-theory in $M$ along which there is no $H_3$ flux. 

With this formulation of generalized global symmetry via twisted K-theory, the matching across T-duality has exactly been developed into a well-studied subject known as topological T-duality, which has been investigated with considerable generality and mathematical rigor. See \cite{Bouwknegt:2003vb,Bouwknegt:2003wp,Bunke:2005sn} for some early works on topological T-duality, and \cite{rosenberg2009topology} for a pedagogical review. Physically, the $B_2$ field is part of the string theory spacetime, so we stick to the mathematical object of twisted K-theory even for a trivial twist class.

\subsection{Examples}

We proceed by computing the even-form and odd-form global symmetries from the twisted K-theory group of $\p X$ in some 6D examples.

\paragraph{6D (2,0) SCFTs}

A 6D (2,0) SCFT can be constructed by compactifying type IIB on $\bbC^2/\Gamma$ with $\Gamma \subset SU(2)$ a finite subgroup. The K-theory of the internal manifold has been computed as (see, e.g., \cite{GarciaEtxebarria:2019caf}):
\begin{equation}
    K^0(S^3/\Gamma) = \bbZ \oplus \text{Ab}(\Gamma) \quad K^1(S^3/\Gamma) = \bbZ,
\end{equation}
where $\Gamma \subset SU(2)$ is the ADE-type finite subgroup of $SU(2)$. For type IIA, on the other hand, the A, D, and E families require a case-by-case analysis, and we only discuss A-type constructed by $k$ NS5 branes in the main text\footnote{See Supplemental Material for the study of D-type (2,0) from type IIA and discussion about E-type (2,0).}. There, the IIA charge group from twisted K-theory is $K^1_{k \xi_3}(S^3) = \bbZ_k$, which does not contain the $\bbZ$ factor as in the ordinary K-theory from IIB side. Indeed, this $\bbZ$ mismatch is removed only if we consider trivially twisted K-theory $K_{0\xi_3}^*(S^3/\bbZ_k)$ for IIB (the computations can be found in \cite{Maldacena:2001xj}) as stated above. As can be seen in both cases, the K-theoretic defect group acts on object that are labeled by $K^0(M)$. In terms of ordinary cohomology elements, $K^0(M)$ encodes elements that have even dimensions, where an interplay between different dimensions is generically expected. %\phy{Maybe more comments on the physical meaning of such interplay, as a ... . Does the case of IIA/IIB on $\bbC^3/\Gamma$ orbifold without D-brane probe give any well-behaved theory?}

Specifically, the relevant twisted K-theoretic symmetries are given by:
\begin{equation}
\begin{array}{cccc}
%             &           IIB     &           IIA         &         \\
    K^0(M) &: K_{0\xi_3}^0(S^3/\bbZ_k) &= K^1_{k \xi_3}(S^3) &= \bbZ_k, \\
    K^1(M) &: K_{0\xi_3}^1(S^3/\bbZ_k) &= K^0_{k \xi_3}(S^3) &= 0.
\end{array}
\end{equation}
See \cite{atiyah2004twisted} for a review of the definition of twisted K-theory. \footnote{We reminder the reader that the twisted cocycle condition for a $\bbZ$-valued twist class has to be defined using the analytic approach in terms of Fredholm operators, so we refrain from introducing it in the main text.} We see that twisted $K^0$ and $K^1$ of $X$ are exchanged under T-duality, so the even-form (resp., odd-form) symmetry in $M$ stays the same. Restricting to the 2-form sector would reproduce the result in \cite{DelZotto:2015isa}, but the K-theoretic symmetry also involves the 0-form and the 4-form sector.

\paragraph{6D (2,0) SCFTs: Worldsheet Perspective}

The analysis of D-branes on $su(2)$ Wess-Zumino-Witten (WZW) model with ADE-type modular invariants can be found in, e.g., \cite{Gaberdiel:2004yn}. In our context, the relevant model should be thought of as the asymptotic boundary on type IIA side, and the resulting charge group for A-brane on $S^3$ with $H_3$-flux is computed from the boundary state formalism of the worldsheet WZW model. The result matches with the expectation from twisted K-theory, which coincides with the center of the ADE Lie group. \footnote{We summarize the details of the worldsheet analysis in the Supplemental Material \ref{apdx:6D_DEtype} for the convenience of the readers.}

This statement is in line with \cite{Kaidi:2024wio,Heckman:2024obe}, the study of gauged generalized symmetries in string spacetime from the worldsheet perspective. Recall that the global symmetry on the string worldsheet with boundary should be the gauge symmetry in the target manifold. If we instead view the target space as the \textit{asymptotic boundary} of the internal manifold, then we will get the twisted K-theoretic global symmetry in the external spacetime dimensions. The author hopes to investigate this delicate match in more generality in the future.

\paragraph{6D (1,0) LST from type II}

Now we go down to 8 supercharges. Some 6D (1,0) theories can be constructed from heterotic string on a $\bbC^2/\Gamma$ orbifold with a suitable flat connection of gauge bundles at infinity \cite{Aspinwall:1997ye}, for which we would need to go beyond T-duality. To postpone such an ambitious move, we focus on theories that thus do not admit heterotic or type I constructions, such as 6D (1,0) theories with non-trivial defect groups. 

To get a relatively rich set of generalized symmetries, we examine 6D (1,0) Little String Theories (LSTs) \cite{Seiberg:1997zk}. LSTs are non-gravitational theories, but they are also non-local since they admit dynamical strings with finite tension (hence ``little string"). Such LSTs can be thought of as intermediate theories between QFTs and quantum theories of gravity. As a remark, the concepts of relative LSTs versus absolute LSTs has not been well-formulated, since LSTs are not local QFTs. Nonetheless, we can still meaningfully discuss the twisted K-theoretic defect group of LSTs, and track their behavior across T-duality. In addition to the even-form (``2-form") symmetries that are more common for 6D theories, LSTs can have discrete ``1-form symmetries" as identified in \cite{Apruzzi:2020zot}. 

We focus on LSTs constructed from IIA on a $\bbC^2/\bbZ_{k_1}$ singularity with $k_2$ NS5 branes. The T-dual construction of the same set of LSTs is given by IIB on $\bbC^2/\bbZ_{k_2}$ with $k_1$ NS5 branes, note again the exchange $k_1 \leftrightarrow k_2$. We express the LST with tensor branch notation introduced in \cite{Bhardwaj:2015oru} as:
\begin{equation}
    \slash\slash \overset{\mathfrak{su}(k_1)}{2}\ \dots  \overset{\mathfrak{su}(k_1)}{2}  \slash\slash \quad  (k_2\text{ tensors}).
\end{equation}
Suppressing the details of string geometry, ``$// 2\ 2\ \dots 2//$" encodes the Dirac pairing of $k_2$ tensor multiplets in 6D that form an A-type affine Dynkin diagram, while $\ksu(k_1)$ stands for the gauge algebra paired with each tensor multiplet on the tensor branch. The match of the twisted K-theory symmetry can be seen on both sides as:
\begin{small}
\begin{align}
\begin{array}{cccc}
            &           IIB     &           IIA         &         \\
    K^0(M) &: K^0_{k_2 \xi_3}(S^3/\bbZ_{k_1}) &= K^1_{k_1 \xi_3}(S^3/\bbZ_{k_2}) &= \bbZ_{k_1} \\
    K^1(M) &: K^1_{k_2 \xi_3}(S^3/\bbZ_{k_1}) &= K^0_{k_1 \xi_3}(S^3/\bbZ_{k_2}) &= \bbZ_{k_2}.
\end{array}
\end{align}
\end{small}
We get a new odd-form symmetry $\bbZ_{k_2}$, which is an uplift of the 1-form symmetry identified in \cite{Apruzzi:2020zot}. The K-theoretic symmetries are indeed consistent under T-duality.

%\textbf{4D orbifolds} \hyz{Delete if we have nothing to add.}

\section{Symmetry Operator \& Charged Object}\label{sec:TachyonCondensation}

Having presented the description of generalized global symmetries by twisted K-theory, we now turn to their physical interpretations. Compared to generalized symmetries by ordinary cohomology, K-theory mixes up $p_i$ form and $p_j$-form symmetries for $p_i - p_j$ even. In such a theory, the notion of $p$-form symmetries no longer makes sense for individual $p$, in that there is a single group controlling the spacetime symmetry operators of all odd (or even) dimensionalities. In this sense, it can be viewed as a specific type of higher-group symmetries, with the specific form given by its string-theoretic construction.

\paragraph{K-theoretic Symmetry as Equivalence Classes of Symmetry Operators}

Indeed, D$p$-branes for different $p$ can wrap the same boundary homology cycle. They become symmetry operators of different dimensionality charged under the same (even- or odd-form) symmetry in the non-compact spacetime. This is consistent with the fact that individual components of RR potentials, $C_i$, are not independent physical objects, but only the K-theory class represented by their formal sums are gauge-invariant objects (see the $\bbR\bbP^7$ example in \cite{Moore:1999gb}, which will show up again as the asymptotic boundary of $\bbC^4/\bbZ_2$ below). Therefore, the symmetry operators that one obtains by integrating these $C_i$ over submanifolds of the spacetime are not independent objects, either. In particular, different components of the RR fields $C_0, C_2, \dots$ in the Wess-Zumino action on the D-brane worldvolume $WZ(X) = 2\pi i \int_X e^{F_2 - B_2} \sqrt{\widehat{A}} (C_0 + C_2 + \cdots)$ should not be thought of as independent objects, but instead as objects that are correlated according to the Chern character from a K-theory class $x \in K^0(X)$ to $ch(x) \in H^{\text{even}}(X, \bbQ)$ (we took type IIA as an example). In addition, \cite{GarciaEtxebarria:2019caf} analyzed the K-theoretic RR fluxes and RR charges in IIB compactification, with 6d theories as a family of examples. This suggests that the notions of defect group, absolute theories, and polarizations, should all be phrased in terms of such K-theoretic symmetry of the spacetime manifold, which is in line with our proposal.

\paragraph{6D (2,0) SCFTs} From IIB, the ``2-form $\bbZ_N$ symmetry" was originally constructed by D3 branes wrapping on relative two cycles of $\bbC^2/\Gamma$ \cite{DelZotto:2015isa}. However, on the same cycle, we can also wrap D1 branes or D5 branes, which will result in 0-dimensional or 4-dimensional (in addition to the 2-dimensional) non-dynamical defects in the spacetime. The apparently independent symmetries $\bbZ_N^{(0)}, \bbZ_N^{(2)}, \bbZ_N^{(4)}$ actually uplifts to a single $\bbZ_N$-valued even-form K-theoretic symmetry. Wrapping D5 brane on the resolution divisor of $\bbC^2/\Gamma$ should produce for us Gukov-Witten operators \cite{Gukov:2008sn} that are charged under the same $\bbZ_N^{(even)}$ as the surface defects in \cite{DelZotto:2015isa}.

From IIA side, due to the presence of $k$ units of $H_3$ flux, the combination of D$p$-brane will detect twisted relative cycles $\gamma_{odd} \in H^*_{k\xi_3}(S^3)$, as specified in equation (\ref{eqn:H_twist_S3}). But these twisted cycles are already living in a $\bbZ_2$-graded cohomology, so it does not have a $\bbZ$-valued degree. If we take the wrapped D3-brane configuration and attempt to T-dualize it to the IIA side, then we will get either a D2 brane or D4 brane on the IIA side depending on the specific duality circle we pick, i.e., a double T-duality would exchange D2 with D4. These apparent difficulties in using wrapped D-branes all support the proposal of describing the generalized symmetries by twisted K-theory, which is then naturally compatible with T-duality.

\section{IIB on $\mathbb{C}^3$ and $\mathbb{C}^4$ orbifolds and their dual descriptions}\label{sec:c3c4}

In this section, we expand upon the previous discussion to cover more string compactifications on a higher-dimensional internal manifold, which we take to be quotients of $\bbC^3$ and $\bbC^4$, resulting in a lower-dimensional QFT. As we will see,  in the case of $\bbC^3$ orbifold with a freely acting quotient group that preserves the supersymmetry, the K-theory of the asymptotic boundary $K^*(S^5/\Gamma)$ is simply given by a graded sum of the cohomology ring
\begin{equation}
    K^{i}(S^5/\Gamma) = \bigoplus_{j = i + 2n} H^j(S^5/\Gamma), \label{eqn:noextension}
\end{equation}
without any non-trivial extension. However, there exist supersymmetric $\bbC^4$ orbifolds with isolated singularities whose asymptotic boundary $S^7/\Gamma$ does not obey the above identity. Namely, the K-theory description in such examples gives a global symmetry that is obtained from the cohomology description via a non-trivial extension, i.e., different from (\ref{eqn:noextension}). This predicts 0-form defect groups which are not detectable using cohomology alone. 

In the end, We further review their brane duals given by brane tiling (dimer models) \cite{Hanany:2005ve,Franco:2005rj} dual to IIB on $\bbC^3$ orbifolds via double T-duality, and brane brick model \cite{Franco:2015tya,Franco:2016qxh} dual to IIB on $\bbC^4$ orbifolds via triple T-duality. These cases turns out to provide cases that goes beyond the current framework of topological T-duality.

\subsection{$\mathbb{C}^3$ orbifold}

We first determine the K-theoretic symmetries in $\bbC^3$ orbifold computations. For simplicity, we focus on cases of quotienting by a cyclic group $\bbZ_k$ for the moment.

\textbf{Free action case} For $S^5/\bbZ_k$ (that is the boundary of $\bbC^3/\bbZ_k$) with the $\bbZ_k$ acting freely, the cohomology is given by $H^*(S^5/\bbZ_k) = \{\bbZ, 0, \bbZ_k, 0, \bbZ_k, \bbZ\}$ which follows from explicit computation. As for K-theory, we have $K^1(S^5/\bbZ_k) = \bbZ$ (which comes from the top-degree cohomology) and $K^0(S^5/\bbZ_k) = \bbZ \oplus T$, where $T$ is a torsional group of order $k^2$, but a priori, the precise form of $T$ depends on $k$ and the specific action using the Atiyah-Hirzebruch spectral sequence (e.g., see \cite{Maldacena:2001xj} for an example of such computation). In case such an extension happens, it would means that the symmetry operators obtained by \textit{$Dp$ branes wrapped on $q$-cycles in $\p X$} with different $(p, q)$ pair no longer give rise to independent topological operators, but their symmetry group are organized into a large group in a ways that is precisely controlled by such an extension.

We first consider supersymmetric orbifolds, namely cases where $\bbZ_k$ is a subgroup of $SU(3)$. For supersymmetric quotients, we need the quotient to be a subgroup of $SU(3)$, namely a quotient of the form $\frac{2\pi i}{N}(k_1, k_2, k_3)$ (which we denote as $\bbZ_N^{(k_1, k_2, k_3)}$ for compactness) need to satisfy the condition that
\begin{equation}
    k_1 + k_2 + k_3 = 0 \ \ \mod N.
\end{equation}
As we will show below, non-trivial extension happens at prime two, which is only applicable for $N$ even. However, having isolated singularities would imply $(k_i, N) = 1$ for $i = 1, 2, 3$, which then means that all $i$'s are odd and thus violating the supersymmetric condition. 

An algorithm is provided in \cite{hong2003quantum} for computing K-theory of weighted lens space of $S^{2n-1}/\bbZ_N^{(m_1, m_2, ..., m_n)}$ with free action, i.e., as long as all the weights $m_i$ are coprime with $N$. \footnote{In fact, the algorithm covers the K-theory of \textit{quantum} weighted lens space parametrized by $q \in (0, 1]$, but it reduces to ordinary weighted lens space upon taking $q = 1$.} Using their approach, we can obtain the K-theory of $S^5/\bbZ_N^{(m_1, m_2, m_3)}$ for arbitrary $m_1, m_2, m_3$ as follows:
\begin{equation}
     K^0(S^5/\bbZ_N^{(m_1, m_2, m_3)}) = \left\{ \begin{array}{cc} \bbZ \oplus \bbZ_N^2 & (N \text{ odd }) \\ \bbZ \oplus \bbZ_{2N} \oplus \bbZ_{N/2} &  (N \text{ even })\end{array}\right., \quad K^1(S^5/\bbZ_N^{(m_1, m_2, m_3)}) = \bbZ
\end{equation}
The special case for $m_2 = m_3$ is covered in Proposition 2.3 of \cite{hong2003quantum}. The general case after lifting this condition can be obtained via generalizing the original analysis. \footnote{The general statement is that 
\begin{equation}
K_0(S^5/\bbZ_N) = \text{Coker}(\Phi), K_1 = \text{Ker}(\Phi),
\end{equation}
where $\Phi$ is a certain graph adjacency matrix which then depends on $N, m_1, m_2, m_3$. Lifting the condition of $m_2 = m_3$ would only introduce slight more complication in solving the linear equations necessary for computing $\text{Coker}(\Phi)$.}

From this result, we learn that non-trivial extension in K-theory may only happen for $N$ even. But then, the condition of $(m_i, N) = 1$ for $i = 1, 2, 3$ would make $m_1 + m_2 + m_3 \neq 0 \mod N$. So we conclude that, for supersymmetric orbifolds of the form $S^{5}/\bbZ_N$, we always have $K_0(S^{5}/\bbZ_N) = \bbZ \oplus \bbZ_p^2$ and $K_1(S^{5}/\bbZ_N) = \bbZ$, and no non-trivial extension relative to ordinary cohomology takes places.

For example, if we compactify type IIB string theory on $\bbC^3/\bbZ_p$ orbifold that are supersymmetic and isolated, we would then get a 4D $\calN = 2$ theory. Then the K-theoretic symmetry would juxtapost the $\bbZ_p$ 0-form symmetry with the $\bbZ_p$ 2-form symmetry.

However, the K theory would only exhibit non-trivial difference from cohomology if at least one of the following conditions applies. We briefly discuss the open questions for each case.
\begin{itemize}
    \item The orbifold is non-supersymmetric, i.e., $m_1 + m_2 + m_3 \neq 0 \mod N$, such as $\bbC^3/\bbZ_2^{(1, 1, 1)}$. In such cases, the vacua would decay via tachyon condensation into a lower-dimensional orbifold. The higher-form symmetries in such orbifolds are studied in \cite{Braeger:2024jcj}.
    \item The orbifold is an extended singularity, such as $\bbC^3/\bbZ_4^{(1, 1, 2)}$ where twice of the generator fixes  the $z=0$ complex plane. In this case, compactification of M-theory has been found to admit 2-Group symmetries in \cite{DelZotto:2022fnw,DelZotto:2022joo,Cvetic:2022imb}, which reduces to a symmetry extension for type IIA on such singularities. It would be very interesting to characterize the K-theoretic version of 2-group symmetries, for which we hope to return in the near future. In this sense, the computation in \cite{hong2003quantum} resolves the extended singularity, so it would be already interesting to understand the meaning of quantum K-theory in encoding the reduction the 2-group symmetry produced by M-theory compactification down to IIA on an $S^1$.
\end{itemize}

\subsection{$\mathbb{C}^4$ orbifolds}

Type II string theory compactified on $\bbC^4/\Gamma$ orbifold with D1 brane probe would generically preserve two supercharges, that is $\calN = (2, 0)$ for IIB and $\calN = (1, 1)$ for IIA. As long as the quotient has no fixed loci on the boundary $S^7/\Gamma$, we can similarly use the prescription of \cite{hong2003quantum} to determine the K-theoretic symmetry of the compactified theory. Now, however, there are indeed \textit{supersymmetric} examples where the K-theory admit non-trivial extension relative to cohomology that differs from (\ref{eqn:noextension}). \footnote{During our revision, we became aware of the upcoming work \cite{CHTWZ_WIP} which also covers these cases, focusing the field-theoretic interpretation of the K-theory symmetry in terms of topological operators and charged states and the difference from the description using ordinary cohomology. We are grateful to the authors for helpful correspon}

\textbf{$\bbC^4/\bbZ_2$ example} It has been known in the string theory literature that $\bbR \bbP^n$ has a K-theory that is different from cohomology. For our case, we need $\bbR\bbP^{2m-1} = \p \bbC^{m}/\bbZ_2$, which can be found in the math textbook \cite{karoubi2009k} by Karoubi. The case for $m = 4$ appeared in the discussions of \cite{Moore:1999gb}. Physically, it means that the K-theory will be different from cohomology with discussing the physics of RR field in type II string theory. In particular, we have $K^*(S^7/\bbZ_2) = \{\bbZ + \bbZ_8, \bbZ\}$ while $H(S^7/\bbZ_2) = \{\bbZ, 0, \bbZ_2, 0, \bbZ_2, 0, \bbZ_2, \bbZ\}$, where the $\bbZ_8$ can be viewed as an non-trivial extension of the three $\bbZ_2$ factors at degree $2, 4, 6$, respectively. Physically, in an IIB reduction, the single $\bbZ_8$ would come from an extension of three $\bbZ_2$'s, each detected by D1 on 1-cycles in $\bbR\bbP^7$ as the asymptotic boundary, D3 on 3-cycles, and D5 on 5-cycles.

In such a case, the K-theoretic defect group is $\bbZ_8$ whose order is not a complete square, so it does not admit a Lagrangian subgroup. Physically, this means that we would not have any choice of boundary condition that lifts the flux non-commutativity and make the theory absolute.

\textbf{$\bbC^4/\bbZ_4$ example} Similarly, we consider the space of $\bbC^4/\bbZ_4^{(1, 1, 1, 1)}$ where the quotient $\bbZ_4$ acts by a 
$K(\bbC^4/\bbZ_4^{(1, 1, 1, 1)})$. We have the K-theory of its boundary $S^7/\bbZ_4^{(1, 1, 1, 1)} = \p \bbC^4/\bbZ_4^{(1, 1, 1, 1)}$ given by  
\begin{equation}
    K(S^7/\bbZ_4^{(1, 1, 1, 1)}) = \bbZ_{16} \oplus \bbZ_2^2, 
\end{equation}
while its cohomology is given by $H^*(S^7/\bbZ_4) = \{\bbZ, 0, \bbZ_4, 0, \bbZ_4, 0, \bbZ_4,  \bbZ\}$, the computation can be done either from a more general approach following \cite{hong2003quantum}, or it can be found specifically in \cite{kobayashi1971k}. 

In terms of an absolute theory, we would need to pick a polarization (i.e., a Lagrangian subgroup) of the defect group $L \subset \bbD$, so that the quotient group $\bbD/L$ gives the symmetry in the 2D theory. For these $\bbC^4$ orbifold case, the symmetry we get would reduce to 0-form symmetries. In these cases, a defect group of $\bbZ_4^3$ would only allow us to have a Lagrangian subgroup of $\bbZ_2^3$ but not $\bbZ_2 \times \bbZ_4$. To see this, we notice that a generic bilinear form on $\bbZ_4^3$ can be taken modulo change of basis to be 
\begin{equation}
    \frac{1}{4}(a_1 b_1 + a_2 b_2 + a_3 b_3), \quad (a_1, a_2, a_3), (b_1, b_2, b_3) \in \bbZ_4^3.
\end{equation}
and a Lagrangian subgroup should be such that any pair of elements $\mathbf{a}, \mathbf{b}$ has a vanishing bilinear form. One can see that a generator of $\bbZ_4$ has trivial bilinear product with itself, thus the only choice of $L$ is $\bbZ_2^3$, giving a 0-form symmetry equal to $\bbZ_2^3$. Physically, the topological lines corresponding to the three generators are given by D1, D3, D5 brane wrapping 1, 3, 5 cycles, respectively. On the other hand, by using K-theory, we would get $\bbD = \bbZ_{16} \oplus \bbZ_2$, with a bilinear form $\frac{1}{16} a_1 a_2 + \frac{1}{2} b_1 b_2$. Now, we are indeed allowed to pick a polarization of $L = \bbZ_4 + \bbZ_2$, thus giving a 0-form symmetry of $\bbD/\bbL \cong \bbZ_4 \oplus \bbZ_2$.

Via the Atiyah-Hirzebruch Spectral Sequence (AHSS) computation, one can in principle extract how these generators of K-theory reorganizes the generators of cohomology at degree 2, 4, and 6 via the extension problem on the $E_\infty$ page of the spectral sequence. Even though the specific computation done in \cite{kobayashi1971k} and \cite{hong2003quantum} does not make use of AHSS, it would still prove useful for us to discuss this approach, due to its physical interpretation as discussed in \cite{Maldacena:2001xj}. Here, after stabilizing to the $E_\infty$ page, there remains a step of extension:
\begin{equation}
    K^0(S^7/\bbZ_4) = \bbZ_{16} \oplus \bbZ_2^2 := \mathrm{Gr}_0 \supset \mathrm{Gr}_1 \supset \mathrm{Gr}_2 \supset \mathrm{Gr}_3 = 1.
\end{equation}
Such that $\mathrm{Gr}_i/\mathrm{Gr}_{i+1} = \bbZ_4$. Here, the intermediate groups are actually unique: 
\begin{equation}
    \mathrm{Gr}_{1} = \bbZ_{8} \oplus \bbZ_2, \quad \mathrm{Gr}_2 = \bbZ_4
\end{equation}
the extension of AHSS is such that  $\mathrm{Gr}_i/\mathrm{Gr}_{i+1} \cong H^{2i+2} = \bbZ_4$, so we know that each of the two $\bbZ_2$ factors in $K^0(S^7/\bbZ_4)$ are only associated with D3 wrapping 3-cycle, and D5 wrapping 5-cycles, respectively. On the other hand, the $\bbZ_{16}$ generator spreads out into all three degrees, encoding the entire $\bbZ_4$ of D1 wrapping 1-cycle and a $\bbZ_2$ for each wrapped brane configurations of (D3, 3-cycle) and (D5, 5-cycle). 

The degree-counting argument goes similarly in the $\bbC^4/\bbZ_2$ case, where the three $\bbZ_2$ extends into a single $\bbZ_8$ in a particular order. The ``finest" $\bbZ_2$ coming out of $\bbZ_8/\bbZ_4$ is associated with D1 on 1-cycle, and the second finest $\bbZ_2$ is associated with D3 on 3-cycle. This is in accordance with the general idea of tachyon condensation: one always starts from higher-dimensional brane-antibrane pairs, which annihilates to give lower-dimensional branes. So for topological lines coming from different wrapped branes, the higher-dimensional wrapped brane charge is extended and refined by the lower-dimensional wrapped brane charge.

\subsection{Discussion: dual brane descriptions}

In the end, we give some brief discussion on extending the topological T-duality discussion to $\bbC^3$ and $\bbC^4$ orbifolds. The main message is that, the T-dual brane configuration will contain NS5 branes with legs in the periodic directions (i.e., $T^2$ or $T^3$), so the asymptotic infinity will still contain these torus, where at places infinitely close to the NS5 branes the $H_3$ flux will diverge. This presents a challenge in formulating the twisted K-theory in such brane configurations. 

\textbf{Brane tilings (dimer models)} \cite{Hanany:2005ve,Franco:2005rj} provide a concrete brane construction of 4D $\mathcal{N}=1$ quiver gauge theories, which are T-duals of D3-branes probing toric Calabi--Yau threefold ($\mathrm{CY}_3$) singularities. Upon compactifying two internal directions into a $T^2$ and performing a T-duality along each direction, the D3 turns into D5 wrapping the $T^2$, and the geometric singularity turns into NS5 branes that divides $T^2$ into ``tiled" pieces. 

% The physical origin lies in Type IIB configurations where NS5-branes form a bipartite, doubly-periodic web on a two-torus $\mathbb{T}^2$, while D5-branes fill the faces of the web. The effective 4D gauge theory arises of D3-branes extended along $(x^{0,1,2,3})$. Via T-duality along the torus directions, one obtains an equivalent geometric engineering of D3-branes probing toric $\mathrm{CY}_3$ cones . This chain of dualities provides a first-principles derivation of the dimer data from string theory.

\begin{table}[h]
\centering
\begin{tabular}{c|cccc|cccc|cc}
\hline
 & $x^0$ & $x^1$ & $x^2$ & $x^3$ & $x^4$ & $x^5$ & $x^6$ & $x^7$ & $x^8$ & $x^9$ \\
\hline
D5 & $\times$ & $\times$ & $\times$ & $\times$ &  & $\times$  &  & $\times$ &  &  \\
NS5 & $\times$ & $\times$ & $\times$ & $\times$ & \multicolumn{4}{|c|}{$\Sigma$} &  &  \\
$M_{3, 1}$ & $\times$ & $\times$ & $\times$ & $\times$ &  &  &  & &  &  \\
$\mathbb{T}^2$ &  &  &  &  &  & $\circ$ &  & $\circ$ &  &  \\
\hline
\end{tabular}
\caption{Worldvolume directions for brane tilings in Type IIB.}
\label{table:tiling}
\end{table}

The specific directions each brane populate is explained in Table \ref{table:tiling}. There we see that, the NS5 would populate a Riemann surface $\Sigma \subset \bbC^2_{(4567)}$, thus having one direction within the torus $T^2_{(57)}$ and the other direction outside. 

These real five-dimensional pairs of asymptotic boundaries are in contrast with the simpler T-dual pair of $S^3/\bbZ_k$ and $S^3$ with $k$-units of $H_3$ flux, where the NS5 brane stays in the bulk without touching the asymptotic boundary. As a concrete example, the case of $\bbC^3/(\bbZ_M^{(1, -1, 0)} \times \bbZ_N^{1, 0, -1})$ is dual to a configuration with $M$ NS5 brane along one direction along $T^2$, a $N$ NS5 branes on another circle (e.g., this follows from degenerate cases given in \cite{Franco:2015tna}). In these cases, there are NS5 branes strectching on the $T^2$ directions.  While in the more rudimentary case of $S^3/\bbZ_N$ dual to $S^3$ with $N$ units of $H_3$ flux, the NS5 brane only sits at the ``bulk $\bbC^2$ but not touching the boundary $S^3$.

Here, the proper way to understand the symmetry via K-theory presumably involves Mayer-Vietoris sequence, isolating the NS5-brane loci asymptotic boundary as fibration of $S^3_i$, one for each stack of coincident NS5 branes. This is reminiscent of the discussion of understanding of 2-group symmetries of M-theory compactified on $\bbC^3$ orbifolds with non-isolated singularities \cite{DelZotto:2022fnw,DelZotto:2022joo,Cvetic:2022imb}. However, the two cases should be different, since the configuration of NS5 branes appearing in the asymptotic boundary is a more general phenomenon, regardless of whether the orbifold side involves an isolated singularity or a non-isolated singularity. We leave such an analysis to future work.

\paragraph{Brane brick models and 2D $(0, 2)$}
Brane brick models provide the higher-dimensional analogue of brane tilings for two-dimensional $(0,2)$ supersymmetric quiver gauge theories realized on D1-branes probing toric Calabi--Yau fourfold ($\mathrm{CY}_4$) singularities. The Type IIB brane construction involves NS5-branes forming a three-dimensionally periodic web on a three-torus $\mathbb{T}^3$, with D5-branes filling three-dimensional ``bricks'' and D1-branes supporting the $(0,2)$ gauge dynamics. As in the dimer case, a chain of T-dualities and mirror symmetry relates this configuration to D1-branes probing toric $\mathrm{CY}_4$ cones, while the mirror geometry is captured by an M5-brane wrapping a holomorphic three-dimensional variety in $(\mathbb{C}^\ast)^3$ \cite{Franco:2015tna,Franco:2016qxh}.

These theories are known to be T-dual to IIB on $\bbC^4$ orbifolds, where we have explained how K-theory points to data which cannot be detected by ordinary cohomology. A specific example of brane brick model $\bbC^4/\bbZ_4$ orbifold is studied in \cite{Franco:2024mxa}. There, it would be interesting to identify the similar extension via K-theory from the quiver side. Recall that back in the $\bbC^3$ orbifold case, the higher-form symmetry data could be read out in \cite{Tian:2021cif} from the McKay quiver via Ito-Reid McKay correspondence \cite{ItoReid}. In this case, it would be interesting to see how the K-theoretic symmetries explained above, which encodes extensions between symmetry operators via wrapped D-brane states, manifest in the dual quiver description associated with the brane brick model. It would further be interesting to explore the implications in the context of McKay correspondence. 

\appendix
\section{6D (2,0) DE-type Theories} \label{apdx:6D_DEtype}

In this appendix, we discuss the situation of K-theoretic symmetries for 6D (2,0) theories of DE-type. We will see that, the D-type case can be covered by introducing an extra $\bbZ_2$-valued twist; whereas the E-type case requires comparing type IIB string with M-theory which we defer to future work.

\paragraph{D-type}

Here, the IIB side is a pure geometry $\bbC^2/\Gamma_{D_N}$ with a non-Abelian quotient by the dihedral group $\Gamma_{D_N} \subset SU(2)$ of $4N-8$ elements, such that the even-form charge lattice of non-dynamical defects in the 6D theory is the $D_N$ weight lattice. Therefore, it is straightforward to determine:
\begin{align}
   \text{Ab}(\Gamma_{D_N}) = \left\{ \begin{array}{cc} \bbZ_2 \oplus \bbZ_2 & N \text{ even}, \\ \bbZ_4 & N \text{ odd}.
   \end{array} \right.
\end{align}

However, the K-theoretic symmetry from IIA side needs another type of twist specified by an element of $H^1(X, \bbZ_2)$. According to section 4.2 of \cite{Hanany:2000fq}, 6D (2,0) SCFT of D-type can be engineered by putting a stack of $k$ NS5 brane on top of a $ON5_A$ brane, an object that implements the combination of $I_4$, an inversion of the transverse directions, and $(-1)^{F_L}$, the worldsheet left-moving fermion parity. The K-theory charge of this configuration was computed in \cite{Braun:2004qg}. Concretely, they computed $^-K_{(k+1) \xi_3}^*(S^3/\bbZ_2)$, the K-theory on $S^3/\bbZ_2 \cong \bbR \bbP^3$ with a $\xi_3$ the generator of $H^3(X, \bbZ)$ and a non-trivial 1-form twist $- \in H^1(S^3/\bbZ_2, \bbZ_2) \cong \bbZ_2 := \{\pm \}$ as:
\begin{align}
    ^-K_{(k+1)\xi_3}^0(S^3/\bbZ_2) &= 0, \\
    ^-K_{(k+1)\xi_3}^1(S^3/\bbZ_2) &= \left \{ \begin{array}{cc} \bbZ_2 \oplus \bbZ_2 & k\text{ even,} \\ \bbZ_4 & k\text{ odd.} \end{array}\right.
\end{align}
Here $k + 1 = \kappa$ describes the D-type theory of type $SO(2k)$, and the ``$+1$" accounts for the fact that the orbifold $IIA/I_4 (-1)^{F_L} := \text{ON5}_A^0$ itself carries one unit of $H_3$ flux, which could be viewed as the combination of $\text{ON5}_A$ (without $H_3$ flux) and an NS5 brane.

\paragraph{E-type}

For decades, it was believed that 6D (2,0) SCFT can only be engineered from type IIB string theory via its compactification on a $\bbC^2/\Gamma_{E_N}$ singularity, where $N = 6, 7, 8$. So we learn that their even-form symmetry is $\bbZ_{m}$ for $\Gamma_{E_n}$ with $m = 9-n$, which coincides with the center of the ADE Lie group.

As said, the type IIA construction of such theories is not available. Nonetheless, one could only construct them from M-theory using the exotic U-fold construction \cite{Garcia-Etxebarria:2016erx} which is inherently strongly coupled and thus non-geometric. In M-theory, the five-dimensional internal manifold can be given as $(\bbC \times T^3)/\bbZ_p$ where $\bbZ_p$ is a non-geometric quotient given by:
\begin{equation}
    \bbZ_p = \bbZ_p^\bbC \cdot \bbZ_p^\rho,  
\end{equation}
where $\bbZ_p^\bbC$ is a $\frac{2\pi}{p}$ rotation on the $\bbC$ complex plane, $\bbZ_{p}^{\rho}$ is an order-$p$ element in the non-geometric part $SL(2, \bbZ)_\rho$ of the U-duality group $SL(2, \bbZ)_\rho \times SL(3, \bbZ) \cong E_{3, 3}(\bbZ)$ acting on $T^3$, whereas the geometric factor $SL(3, \bbZ)$ is not involved in the quotient. Such quotients are only possible when the complexified K\"{a}hler moduli is set fixed, so these configurations are inherently strongly coupled and beyond the reach of perturbative IIA. As stated, we leave the study of charges in M-theory to future work.

\section{Worldsheet Analysis} \label{apdx:WS}

Here we summarize the worldsheet analysis of D-brane charges in various literature in the 6D (2,0) cases in \cite{Gaberdiel:2004yn} for the convenience of the readers. As an illustration, we focus on the analysis of D-type cases. We consider $\ksu(2)_k$ WZW models for even $k$ with D-type non-diagonal modular invariants, also known as $\kso(3)_k$ WZW model. We remark that this only covers the worldsheet of IIA construction of the D-type (2,0) theories. There, the spectrum of this theory in terms of holomorphic and anti-holomorphic states is given by:
\begin{align}
\begin{split}
    \calH = &\bigoplus_{j = 0}^{k/2} \calH_{2j} \otimes \overline{\calH}_{2j} \oplus \bigoplus_{j = 1}^{k/2} \calH_{2j-1} \otimes \overline{\calH}_{k+1-2j} \quad (k/2\ \text{odd}), \\
    \calH = &\bigoplus_{j = 0}^{k/4-1} \left(\calH_{2j} \oplus \calH_{k-2j}\right) \otimes \left( \calH_{2j-1} \oplus \overline{\calH}_{k+1-2j} \right) \\
            &\oplus 2 \calH_{k/2} \otimes \overline{\calH}_{k/2} \quad (k/2\ \text{even}).
\end{split}
\end{align}

The \textit{indices} $\mu, \nu, \dots$ refer to bulk fields that control the bulk modular S-matrix $S_{\mu\nu}$ and fusion coefficients $\calN_{\mu\nu}^{\ \ \ \rho}$.  For concrete Hilbert spaces as given above, the number of bulk indices is given as the number of $\calH_a \otimes \overline{\calH}_a$ (``diagonal entries") after expanding out the tensor product, which then equals to the number of untwisted boundary states. One can count the bulk indices to get the number $n = k/2 + 2$, i.e., there are $n$ (untwisted) boundary states. See \cite{Cardy:2004hm} for a pedagogical treatment of the boundary CFT formalism, including definitions and explicit expressions of these states. 

Here we consider the standard situation that the boundary corresponds to the regular representation, so the set of index of such boundary states (usually denoted as $a, b, \dots$) coincides with the set of bulk indices. Then the boundary state charges are labeled by nodes on $D_n$-type Dynkin diagram, where the $i$-th node is associated with a charge $q_i$. These charges need to satisfy the charge equations modulo 2 \cite{Fredenhagen:2000ei}:
\begin{align}
\begin{split}
    & 2q_i = q_{i-1} + q_{i+1} (1 < i < n-2), \\
    & 2q_1 = q_2,\ 2q_{n-1} = 2q_{n} = q_{n-2},\\
    & 2q_{n-2} = q_{n-3} + q_{n-1} + q_{n}
\end{split}
\end{align}
Here $i = 1, \dots, n$ label nodes on a $D_n$ Dynkin diagram, where $i = n-2$ labels the trivalent node whose three adjacent nodes are then labeled by $i = n-3,\ n-1,\ n$, respectively. Solving these relations would give us the charge group $K = \bbZ_2 \oplus \bbZ_2$ for $k$ even and $K = \bbZ_4$ for $k$ odd.

The analysis of A-type and E-type proceeds in a completely analogous fashion, where in the $E$-type we need to use the $su(2)_k$ E-type modular invariant partition functions for $k = 10, 16, 28$. The outcomes all coincide with the center of ADE simply-laced Lie group via McKay correspondence.

\acknowledgments

The author thanks Ibrahima Bah, Jiakang Bao, Saghar S. Hosseini, Max H\"ubner, Yuji Tachikawa, Yi-Nan Wang, Xingyang Yu for discussions, and Yuji Tachikawa for helpful comments on an earlier version of the draft. The author thanks Jonathan Heckman, Max H\"{u}bner, Ethan Torres, and Xingyang Yu for collaboration on the role of K-theory in generalized symmetries and geometric engineering, together with Justin Kaidi and Yuji Tachikawa for collaboration on the spacetime implication of generalized symmetries on the string worldsheet. In the end, we are grateful for the comments by the anonymous referee, which have significantly improved the content and the presentation of our paper. The author is supported in part by WPI Initiative, MEXT, Japan at Kavli IPMU, the University of Tokyo. 

% Bibliography

%% [A] Recommended: using JHEP.bst file
%% \bibliographystyle{JHEP}
%% \bibliography{biblio.bib}

%% or
%% [B] Manual formatting (see below)
%% (i) We suggest to always provide author, title and journal data or doi:
%% in short all the informations that clearly identify a document.
%% (ii) please avoid comments such as "For a review'', "For some examples",
%% "and references therein" or move them in the text. In general, please leave only references in the bibliography and move all
%% accessory text in footnotes.
%% (iii) Also, please have only one work for each \bibitem.

\appendix

\bibliographystyle{jhep}
\bibliography{T.bib}

\end{document}